\documentclass[11pt,twoside]{article}

\usepackage{asp2006}
\usepackage{epsf}
\usepackage{psfig}
\usepackage{lscape}

\begin{document}
\title{A multiwavelength study of a young, Z-shaped, FR I radio galaxy NGC3801}   
\author{Ananda Hota,$^1$ Jeremy Lim,$^1$ Youichi Ohyama,$^1$ 
        D. J. Saikia,$^2$ \\ Dinh-V-Trung,$^1$ and J. H. Croston$^3$}   
\affil{$^1$ASIAA, P.O. Box 23-141, Taipei 106, Taiwan \\ 
  $^2$National Centre for Radio Astrophysics, TIFR, Pune 411 007, India \\ 
  $^3$School of Physics, Astronomy, and Mathematics, University of Hertfordshire, Hatfield AL10 9AB, U.K.}

\begin{abstract} 
We present preliminary results from a multi-wavelength study of a merger candidate, NGC3801,
hosting a young FR I radio galaxy, with a Z-shaped structure. Analysing archival data from the VLA, 
we find two H{\sc i} emission blobs on either side of the host galaxy, suggesting a 30 kpc sized rotating 
gas disk aligned with stellar rotation, but rotating significantly faster than the stars. 
Broad, faint, blue-shifted absorption wing and an H{\sc i} absorption clump associated with 
the shocked shell around the eastern lobe are also seen, possibly due to an jet-driven outflow. 
While 8.0 $\mu$m dust and PAH emission, from {\it Spitzer} and near and far UV emission from {\it GALEX} 
is seen on a large scale in an S-shape, partially coinciding with the H{\sc i} emission blobs, it reveals a $\sim$2 kpc 
radius ring-like, dusty, starforming structure in the nuclear region, orthogonal to the radio jet axis. 
Its similarities with Kinematically Decoupled Core galaxies and other evidences have been argued for 
a merger origin of this young, bent jet radio galaxy.
\end{abstract}

\section{NGC3801}
NGC3801 is a nearby E/S0 galaxy at a distance of $\sim$47.9 Mpc, with the body of the galaxy 
being crossed by two main dust features (Heckman et al. 1986; Verdoes Kleijn et al. 1999). 
A warped dust lane lies along the optical minor axis
while patchy dust filaments are seen on the eastern and western halves of the galaxy. 
At brightness levels $\mu_{V}$$\sim$23$-$24 mag arcsec$^{-2}$, the galaxy shows a hysteresis 
loop like structure while at even fainter levels, a boxy isophotal structure is seen (Heckman et al. 1986). 
It hosts a small radio galaxy with an angular size of $\sim$50 arcsec (11 kpc), whose jet axis is almost orthogonal
to the rotation axis of the stellar component or orthogonal to the minor-axis dust lane (Heckman et al. 1985). 
Millimetre-wave observations have helped identify a radio core and clumps of CO(1$-$0) emission suggesting a 
r$\sim$2 kpc circum-nuclear rotating  gas disk orthogonal to stellar rotation and perpendicular 
to the radio jet (Das et al. 2005). 
{\it Chandra} observations reveal shock-heated shells of hot gas
surrounding the radio lobes
(Croston, Kraft, \& Hardcastle 2007).
H{\sc i} observations with the Arecibo telescope show gas in both emission and absorption, but
higher resolution observations are required to determine its distribution and kinematics
(Heckman et al. 1983). We present the first ever imaging study in H{\sc i}, dust, UV and Pa$\alpha$ emission.

\section{Results and Discussion}   
Figure 1 shows the radio continuum image superimposed on an optical DSS blue-band
image in the left panel and the total intensity H{\sc i} image superimposed on the 
8.0 $\mu$m dust/PAH emission image from the {\it Spitzer} on the right panel. The dust emission shows a 
prominent linear feature nearly orthogonal to the jet in the central region (r$\sim$2 kpc).
We found that both the 8.0 $\mu$m dust/PAH emission and UV emission from {\it GALEX}  
show similar $\sim$30 kpc wide S-shaped structure, representing young massive star formation in it.  
Our H{\sc i}-emission study with the VLA shows emission blobs on the eastern 
(mostly red-shifted) and western (blue-shifted) sides, roughly coinciding with the tails of 
the S-shaped structure (Figure 1, right panel). These H{\sc i} results suggest a rotating
gas disk (V$_{circ}$$\sim$280 km s$^{-1}$), with velocities nearly twice than that of
the stars (cf. Heckman et al. 1985). In addition, broad, faint, blue-shifted absorption wing 
and an H{\sc i} absorption clump associated with 
the shocked shell around the eastern lobe are seen, possibly due to jet-driven outflow. 
Due to its similarity with kinematically decoupled cores and other properties, we propose that a 
merger between a gas-rich spiral galaxy and an elliptical galaxy has triggered its AGN activity and has shaped
its stellar, gaseous and radio-jet structures. Detailed stellar population synthesis studies to 
understand its time evolution are in progress.

\begin{figure}
\plottwo{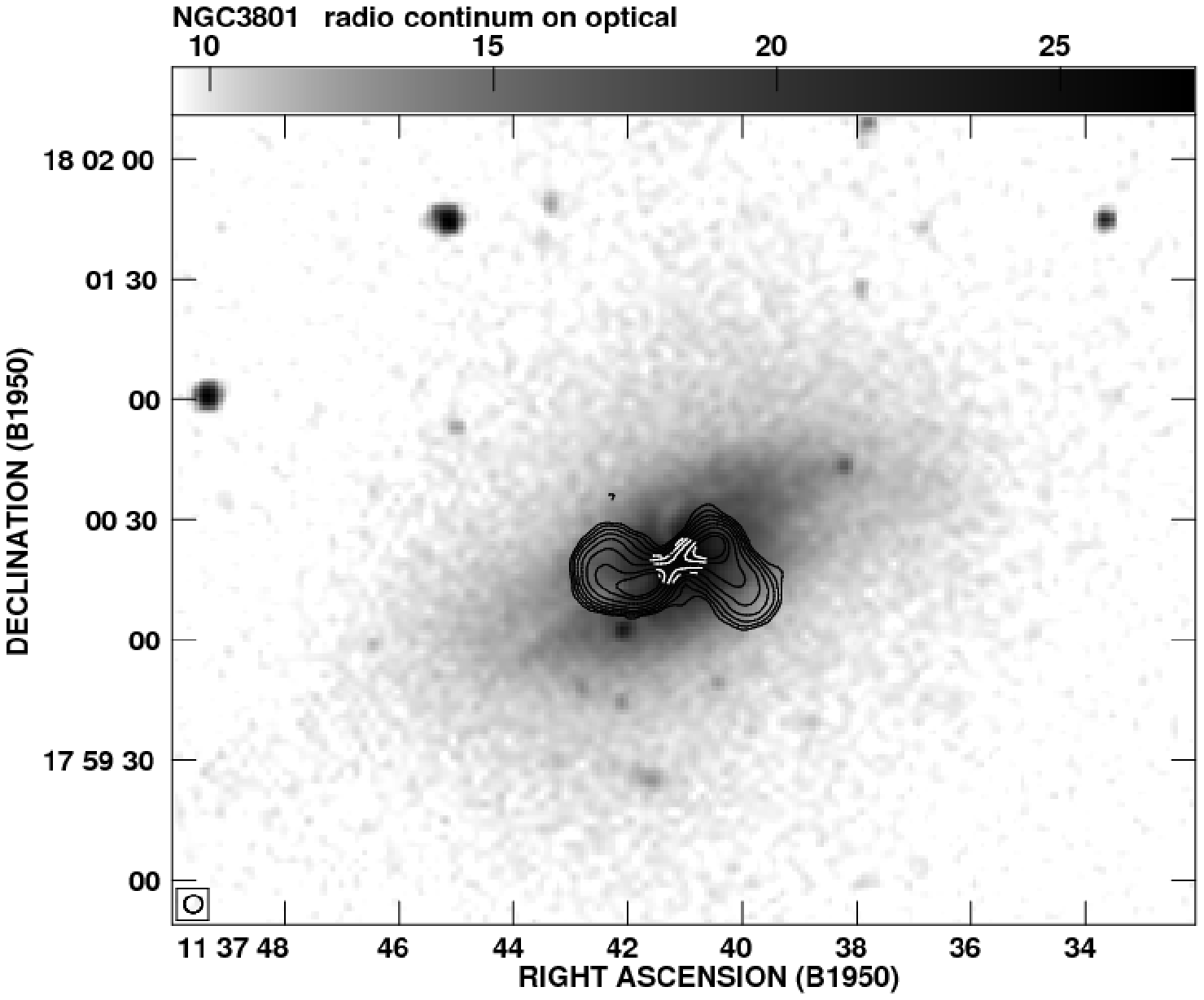}{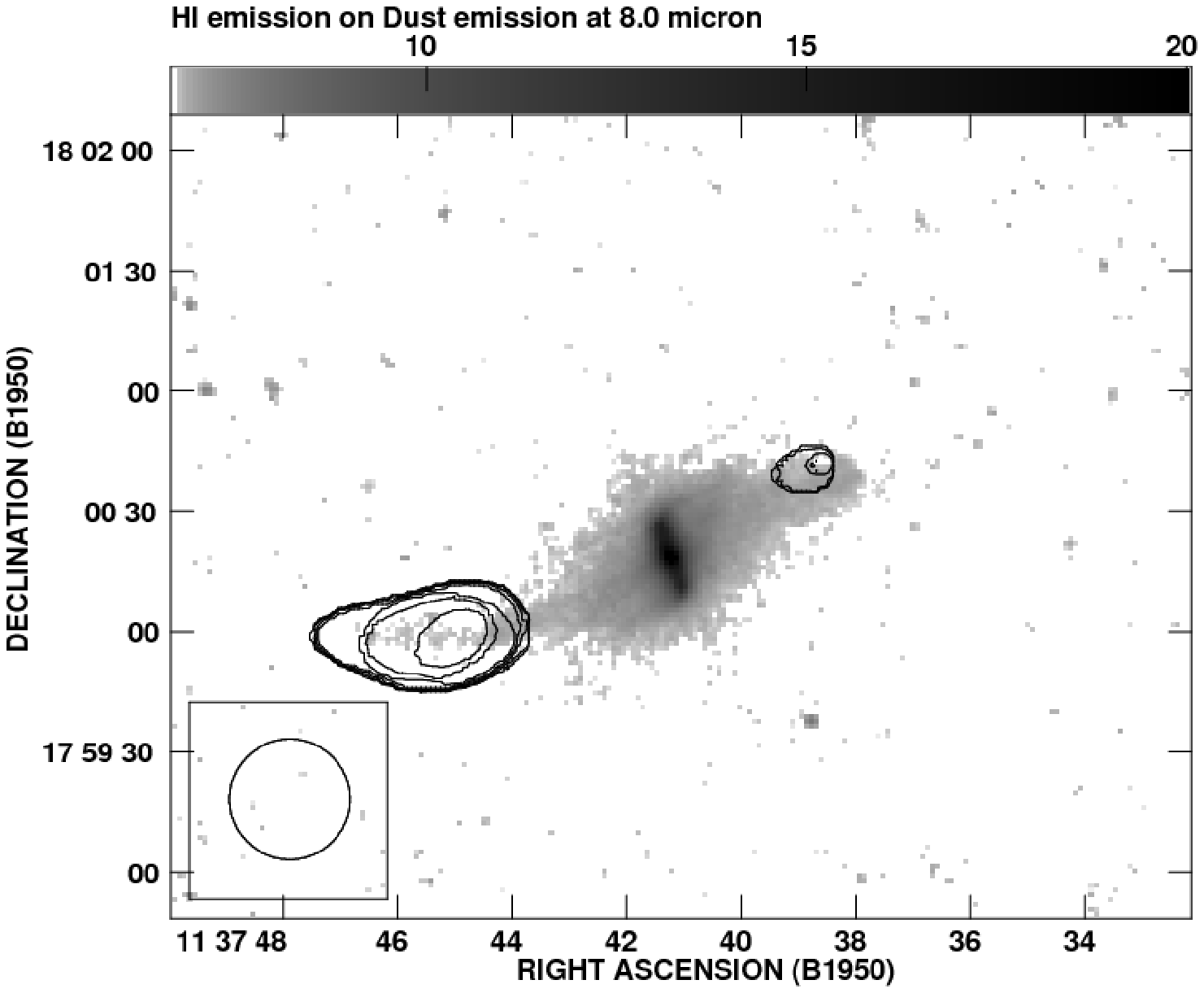}
\caption{Left: 1.4 GHz radio continuum contours superimposed on the optical image. 
Right: Total intensity H{\sc i} emission contours superimposed on dust emission.}
\end{figure}






\begin{thebibliography}{}

\bibitem[]{}Croston, J. H., Kraft, R. P., \& Hardcastle, M. J. 2007, ApJ, 660, 191
\bibitem[]{}Das M. et al.  2005, ApJ, 629, 757
\bibitem[]{}Heckman, T. M., Balick, B., van Breugel, W. J. M., \& Miley, G. K. 1983, AJ, 88, 583
\bibitem[]{}Heckman, T. M., Illingworth, G. D., Miley, G. K., \& van Breugel, W. J. M. 1985, 
            ApJ, 299, 41
\bibitem[]{}Heckman, T. M. et al. 1986, ApJ, 311, 526
\bibitem[]{}Verdoes Kleijn, G. A., Baum, S. A., de Zeeuw, P. T., \& O'Dea, C. P. 1999, AJ, 118, 2592
\end{thebibliography}
\end{document}